\documentclass[notitlepage,pre,aps,onecolumn,amsmath,amssymb,amsfonts,floatfix,superscriptaddress,longbibliography]{revtex4-1}

\usepackage{graphicx}
\usepackage{dcolumn}
\usepackage{bm}
\usepackage{mathrsfs,subcaption}
\usepackage{epstopdf}
\usepackage[thinlines,thiklines]{easybmat}
\usepackage{algorithmic}
\usepackage{appendix}
\usepackage{mathtools} 

\usepackage[space]{grffile}
\usepackage{xcolor}
\usepackage{rotating}
\usepackage{booktabs} 
\usepackage{makecell}

\graphicspath{{./figs/}}

\usepackage[breaklinks=true,colorlinks=true,linkcolor=blue,urlcolor=blue,citecolor=blue]{hyperref}

\usepackage{color}

\begin{document}

\title{Forced symmetry breaking as a mechanism for rogue bursts in a dissipative
  nonlinear dynamical lattice}
\author{P.~Subramanian}  
\affiliation{Department of Mathematics, University of Auckland, 38 Princes Street, Auckland 1010, New Zealand}
\affiliation{Mathematical Institute, University of Oxford, Oxford OX2 6GG, UK}
\author{E.~Knobloch}
\affiliation{Department of Physics, University of California at Berkeley, Berkeley, CA 94720, USA}
\author{P.\,G. Kevrekidis}
\affiliation{Department of Mathematics and Statistics, University of Massachusetts, Amherst MA 01003-4515, USA}


\begin{abstract}
We propose an alternative to the standard mechanisms for the formation of rogue waves in a non-conservative, nonlinear lattice dynamical system. We consider an ODE system that features regular periodic bursting arising from forced symmetry breaking. We then connect such potentially exploding units via a diffusive lattice coupling and investigate the resulting spatio-temporal dynamics for different types of initial conditions (localized or extended). We find that in both cases, particular oscillators undergo extremely fast and large amplitude excursions, resembling a rogue wave burst. Furthermore, the probability distribution of different amplitudes exhibits bimodality, with peaks at both vanishing and very large amplitude. While this phenomenology arises over a range of coupling strengths, for large values thereof the system eventually synchronizes and the above phenomenology is suppressed. We use both distributed (such as a synchronization order parameter) and individual oscillator diagnostics to monitor the dynamics and identify potential precursors to large amplitude excursions. We also examine similar behavior with amplitude-dependent diffusive coupling.

\end{abstract}

\pacs{***}

\maketitle


\section{Introduction}

The formation of rogue waves or spatio-temporal bursts has been investigated by a number of authors~\cite{Kharif2003,Onorato2013} and a number of different mechanisms leading to such waves have been identified. Relevant approaches include completely linear mechanisms (e.g., a superposition of multiple linear waves) or fully nonlinear mechanisms (e.g., modulational instability of a uniform wavetrain)~\cite{Pelinovsky2016,R.Osborne2010}. Some of the original observations stem from the study of North Sea waves~\cite{Mori2002,Haver2004,Walker2004,Adcock2011}, but more recent work has extended similar considerations not only to other areas of the oceans but also to numerous other scientific areas where controlled laboratory experiments are available. These include studies of sloshing of water waves in large tanks~\cite{Chabchoub2011,Chabchoub2012,Adcock2018,Chabchoub2020} and the realm of nonlinear optics~\cite{Solli2007,dudley1,Kibler2016,Tikan}, with some works spanning both areas; see, e.g.,~\cite{Chabchoub2020}. Attempts to establish the rogue wave emergence as a property of a broad class of physics-based models have been reported in ultracold atomic Bose-Einstein condensates~\cite{Charalampidis2018a}, in space plasmas~\cite{Ruderman2010,Sabry2012,Bains2014,Tolba2015} and elsewhere~\cite{Hohmann2010}. 

Arguably, a large portion of the relevant efforts to generate rogue waves has revolved around models of dissipationless dispersive wave propagation, with the nonlinear Schr{\"o}dinger equation~\cite{Sulem1999,Ablowitz2003} (NLS) and its variants providing the central model in this direction. Given that the NLS equation describes the slow evolution of a packet of small amplitude dispersive waves, nonlinear effects are key to the formation of large amplitude rogue burst events. In this context, specific nonlinear solutions, such as the Peregrine soliton~\cite{Peregrine1983}, the periodic in time Kuznetsov-Ma soliton~\cite{Kuznetsov1977,Ma1979} and the periodic in space Akhmediev breather~\cite{Akhmediev1986} have been central to numerous investigations. However, as is well-known, such exactly integrable settings and the analytical solutions available via the inverse scattering transform (and related approaches) are rather sparse, especially so in higher-dimensional settings. It is thus of interest to explore alternative mechanisms of potential broad applicability, including to non-conservative or higher-dimensional systems. Indeed, the recent work of~\cite{PhysRevX.9.011054} captures the relevant motivation well in the statement: ``from the general point of view, the identification of the necessary ingredients for the emergence of rogue waves and extreme events in dissipative systems remains a challenging open problem''.

Here we focus on the fundamentally spatio-temporal nature of the localization of rogue waves and their defining property of ``appearing out of nowhere and disappearing without a trace''~\cite{AKHMEDIEV2009675} but employ an approach that does not rely on the Hamiltonian nature of the problem or on its integrability properties. In dissipative systems sustained rogue wave formation requires the presence of forcing. To capture the essence of rogue waves this forcing must be uniform in space. In this work we leverage the properties of a model problem studying the interaction properties of standing waves in domains of moderate aspect ratio, following~\cite{PhysRevLett.80.5329,MOEHLIS2000263}. This dissipative system satisfies the above requirement where waves arise via a Hopf bifurcation from a trivial state. With Neumann boundary conditions these waves may be of even or odd parity under spatial reflection, and in moderately large domains such waves are nearly degenerate with an approximate interchange symmetry between them, resulting in strong interaction. The state of the system is described by a pair of coupled equations for the two wave amplitudes given by the normal form for a Hopf bifurcation with broken $D_4$ symmetry. This symmetry is a consequence of spatial reflection together with the approximate interchange symmetry between the two modes; the symmetry is broken because the interchange symmetry is not exact, and sample integration of the resulting equations reveals intermittent bursts with amplitudes as large as $10^9$ \cite{moehlis_book}. We refer to these solutions as bursts to emphasize they are localized in time but not in space. The resulting system explains successfully the presence of bursts in experiments including those on binary fluid convection in domains of moderate aspect ratio~\cite{PhysRevA.38.3143,PhysRevLett.80.5329}.

Within the above formulation the spatial degrees of freedom are inevitably slaved to the temporal dynamics of the mode amplitudes. In order to activate spatial degrees of freedom over yet large lengthscales, we consider here a ring of identical, diffusively coupled oscillators of the above type, each of which can generate a large spatially coherent burst (or a sequence of such bursts), depending on parameters. We use the ring geometry in order to generate a spatially periodic system, i.e., to mimic a very large aspect ratio system where local regions remain spatially coherent. The resulting model has the advantage that the dynamics of each oscillator is well understood. In particular, it is understood that the broken $D_4$ symmetry permits the trajectory of each individual oscillator to escape to infinity in finite time and to return from infinity, also in finite time, even when the even and odd standing waves both bifurcate supercritically. On a ring of such oscillators the resulting excitation may be localized at one site, or a small group of adjacent sites, resulting in the generation of a temporally {\it and} spatially localized extreme event, i.e., a rogue wave.

We demonstrate here, via direct numerical simulations, that for weak coupling the above system exhibits rogue events apparently occurring at ``random'' times and at ``random'' locations. Importantly, this is the case even when the parameters characterising the individual oscillators are chosen to generate periodic oscillations only. We offer different diagnostics ranging from the event amplitude distribution function to studying the precursors to the local emergence of large amplitude events, in an effort to obtain diverse perspectives towards a qualitative understanding of the relevant phenomenology. This is followed by an investigation of progressively larger coupling, eventually leading to the synchronization of the entire lattice, as well as couplings that vary over space or time, depending on the site amplitude. For this purpose we leverage various synchronization diagnostics such as the Kuramoto order parameter~\cite{STROGATZ20001}. Our hope is that this initial study will provide motivation for further exploration of alternative mechanisms producing extreme events, and potentially enable their identification in a class of nonlinear lattice systems that can also, in principle, be straightforwardly extended to higher dimensions.

The presentation of our results is structured as follows. In section II, we describe the basic formulation of the model, its parameters and initial conditions at the ordinary differential equation level, first for a single node and subsequently for the diffusively coupled network. In the latter setting, we present the main phenomenology of the system, the diagnostic tools of interest and the resulting findings for different values of the coupling parameter, as well as for amplitude-dependent couplings. Finally, in section III, we summarize our findings and point to some directions for future study. The Appendix presents some further details on the dynamics of the ODEs on a single node of the lattice.


\section{Model}
\label{sec:model}

\subsection{An oscillator with approximate $D_4$ symmetry}

The dynamics near onset in a system exhibiting a Hopf bifurcation with broken $D_4$ symmetry, i.e., on a domain of moderate length, is described by the truncated equations 
\begin{equation}
\dot{z}_{\pm} = [\lambda \pm \triangle \lambda + i(\omega \pm \triangle \omega)]z_{\pm} + A (|z_+|^2 + |z_-|^2) z_{\pm} + B |z_{\pm}|^2 z_{\pm} + C \bar{z}_{\pm} z_{\pm}^2\,.
\label{eqn:temporalsys}
\end{equation}
Here $z_{\pm}$ are the complex amplitudes of the even and odd standing wave modes, the parameters $\triangle \lambda$, $\triangle \omega$ measure the differences in their linear growth rates and onset frequencies, respectively, while $A,B,C$ are complex coefficients. When $\triangle \lambda=\triangle \omega=0$ the two modes follow the same evolution equations and the interchange symmetry between them is exact. Thus, the parameters $\triangle \lambda$, $\triangle \omega$ represent terms that reflect the fact that the two competing standing modes are not in general identical. Equations (\ref{eqn:temporalsys}) assume that in a moderately large domain this effect can be captured at linear order, i.e., via the inclusion of small differences in the growth rates and frequencies of the two competing modes~\cite{PhysRevLett.80.5329}.

New variables ${\cal A},\theta,\phi$ defined in \cite{PhysRevLett.80.5329,MOEHLIS2000263} allow us to completely characterise the solutions of the system with exact $D_4$ symmetry ($\triangle \lambda=\triangle \omega=0$) in terms of three qualitatively different periodic solutions, hereafter $u$, $v$ and $w$. Writing 
\begin{equation}
z_{\pm} = {\cal A}^{\frac{1}{2}} \sin\left(\theta+\frac{\pi}{4}\pm \frac{\pi}{4}\right) \exp\left(i\frac{(\pm\phi+\psi)}{2}\right) \,, \label{eqn:newvars}  
\end{equation}
the $u$ solutions correspond to $\cos\theta=0$, $\cos 2\phi=1$, the $v$ solutions correspond to $\cos\theta=0$, $\cos 2\phi=-1$, and the $w$ solutions correspond to $\sin\theta=0$. A fourth solution, a quasiperiodic state referred to as $qp$, is present in a restricted parameter range. When $\triangle \lambda=\triangle \omega=0$ these states bifurcate simultaneously from the trivial state at $\lambda=0$, and the $u$, $v$, $w$ states then represent invariant subspaces of the system. This is no longer the case when the interchange symmetry is broken, i.e., $\triangle \lambda\ne0$, $\triangle \omega\ne 0$. In this case the $w$ states split into two (the even and odd standing oscillations) and the other states are generated only in secondary bifurcations \cite{MOEHLIS2000263}. We mention that in contrast to the $w$ states, the states $u$ and $v$ represent traveling states~\cite{PhysRevLett.80.5329}. 

When $\triangle \lambda\ne0$, $\triangle \omega\ne 0$ there may be parameter regimes in Eqs.~(\ref{eqn:temporalsys}) with no stable small amplitude oscillations near onset, implying that nontrivial dynamics must take place. In particular, it was found that the solutions in this regime can be attracted to an invariant subspace (the $u/v$ subspace) that extends to {\it infinite} amplitude. Solutions following this subspace reach very high amplitudes, and sample integration of the equations revealed intermittent bursts with amplitudes as large as $10^9$ \cite{moehlis_book}. A detailed study using a rescaled time shows that solutions lying in this invariant subspace are attracted to a saddle point at infinity that is, in turn, connected to a second saddle point at infinity whose stable manifold returns the trajectory back to small amplitude. In terms of the original time the excursion to infinity and back takes a finite time~\cite{PhysRevLett.80.5329,MOEHLIS2000263}. This behavior can be established using the variable $\rho={\cal A}^{-1}$: as ${\cal A}\to\infty$, i.e., as $\rho\to0$, the terms with $\triangle \lambda$, $\triangle \omega$ drop out, and the $D_4$ symmetry becomes exact, allowing a complete description of the dynamics near $\rho=0$.

In the following we consider a parameter combination for which each element in a diffusively coupled ring of such elements is described by Eqs.~(\ref{eqn:temporalsys}) but only displays regular finite amplitude periodic spiking as in Fig.~3(a) of~\cite{PhysRevLett.80.5329}. As a result, the rogue waves we observe are a consequence of the spatial coupling of the elements and {\it not} of their individual behavior at the same parameter values. The corresponding parameter values are
\begin{equation}
\lambda = 0.1,\,\,\triangle \lambda=0.03,\,\,\omega=1,\,\,\triangle\omega=0.02,\,\,A = 1-1.5i,\,\,B=-2.8+5i,\,\,C=1+i
\label{eqn:odeparam}
\end{equation}
and we focus on the generation of extreme events in the resulting {\it lattice system}.


\subsection{Diffusively coupled ring of $N$ nodes}

Our system consists of $N$ identical oscillators on a ring where each of the $N$ nodes is modeled by the dynamics described in Eqs.\,(\ref{eqn:temporalsys}). The oscillators are coupled via nearest-neighbor coupling with diffusion coefficient $K$ as 
\begin{equation}
\dot{z}_{\pm, i} = [\lambda \pm \triangle \lambda + i(\omega + \triangle \omega)]z_{\pm, i} + A (|z_{+, i}|^2 + |z_{-, i}|^2) z_{\pm, i} + B |z_{\pm, i}|^2 z_{\pm, i} + C \bar{z}_{\pm, i} z_{\pm, i}^2 + K \Delta_2 z_{\pm, i}\,.
\label{eqn:temporalsyspde}
\end{equation}
Here, $\Delta_2$ stands for the discrete Laplacian and $i=1,\cdots,N$.

The model allows us to explore the interplay between regular periodic spiking (at each node, in the absence of any diffusion) and the effects of amplitude redistribution via diffusive coupling. In the limit of $K=0$, i.e., with no coupling, we expect to recover regular periodic spiking at each node, albeit with a phase that varies randomly from node to node. In the opposite, diffusion-dominated limit with $K\gg1$, we expect that all nodes display synchronized regular periodic spiking. Hence, as we increase $K$, we expect to see progressive synchronization as revealed, for example, by the Kuramoto order parameter (see, e.g.,~\cite{STROGATZ20001}). The transition between these two regimes and the associated dynamical phenomenologies that it enables are the central topic of interest in the present work. Our principal aim is to determine whether in some intermediate regime, the spiking of a single node, alongside the nearest-neighbor coupling, is able to give rise to a rogue event, offering in this way a viable alternative to the more customary Hamiltonian mechanisms discussed in the introduction. 

We choose the number of oscillators in the ring $N$ in the range where we can define a distributed initial condition (such as a sine wave function) with sufficient resolution over a wavelength and all results in the rest of this paper are for a system of $N=37$ oscillators on a ring. Time simulations evolve the system over the time interval $0<t\le 5000$ and were performed using Matlab's ode23s subroutine with both relative and absolute tolerance being $1\times10^{-5}$. 

\begin{figure}[h!]
\centering{\includegraphics[width=14cm]{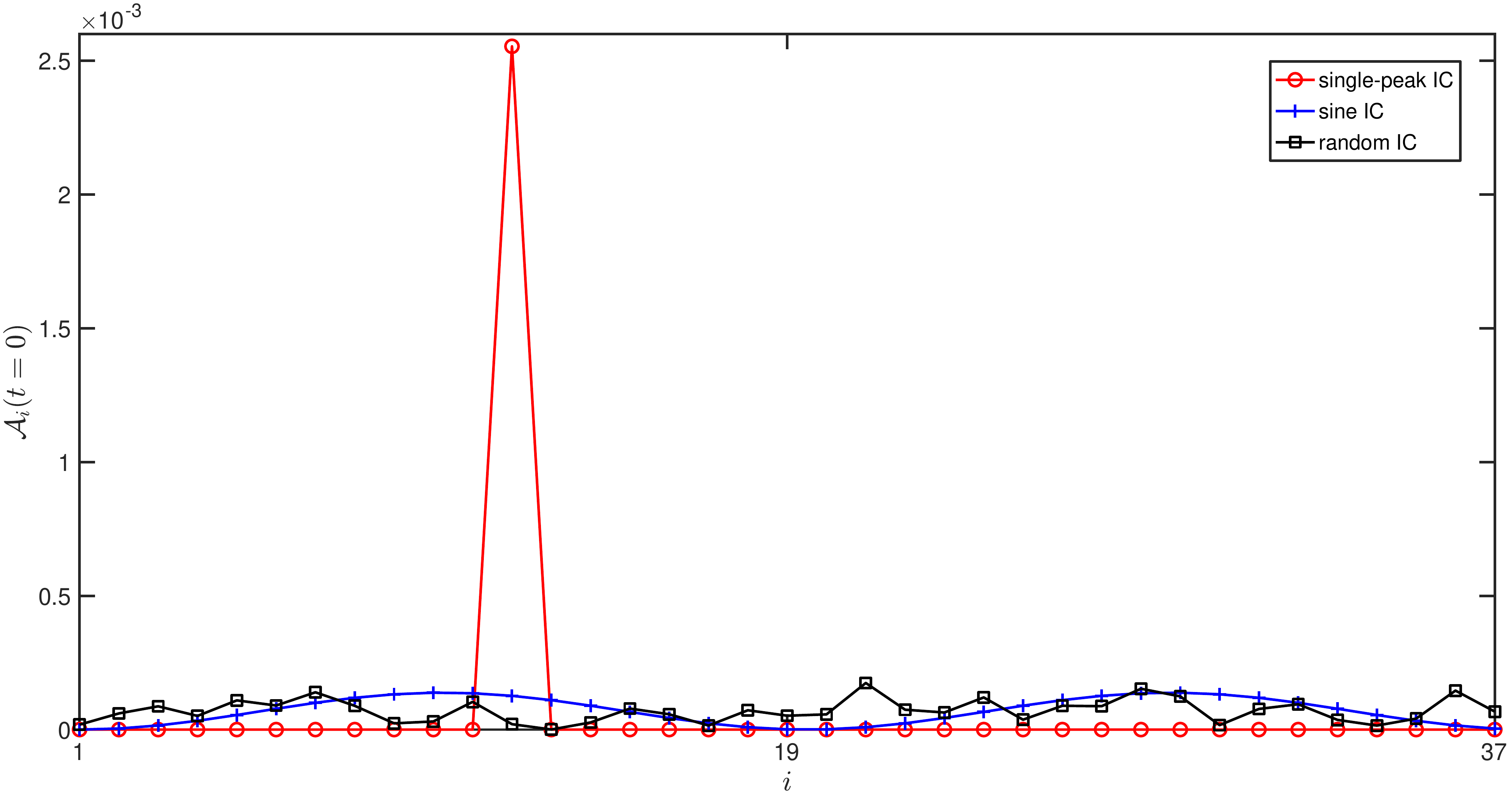}
}
\caption{Three types of initial conditions showing the distribution of initial amplitudes $\mathcal{A}_i(t=0)$ in a system of $N=37$ coupled oscillators. The red line with circle markers shows a single peak initial condition, the blue line with crosses shows a sine wave initial condition and the black line with square markers shows a uniformly distributed random initial condition. In all three cases the
  Riemann sums pertaining to the 3 curves are equal. 
}
\label{fig1:ICs}
\end{figure}

In order to explore the consequence of both a localized/extended initial condition and gradients in the initial condition, we choose three types of initial conditions in our analysis for $z_{\pm}$: a single spike initial condition where $z_{\pm,i}$ is only nonzero at one chosen oscillator, $i=12$, a smoothly distributed initial condition (more specifically a sine wave such that $z_{\pm}$ reaches maximum at $i=10,29$ and minimum at $i=0,19$) and a uniformly distributed random initial condition in (0,1). The amplitude $\mathcal{A}_i$ at the $i$th node at any instant is related to $z_{\pm, i}$ by
\begin{equation}
\mathcal{A}_i = |z_{+, i}|^2 + |z_{-, i}|^2\,;
\end{equation}
our initial conditions are such that the sum of $\mathcal{A}_i$ at $t=0$ over all nodes in the ring is the same. Figure \ref{fig1:ICs} shows a comparison of the three types of initial conditions when $\sum_i \mathcal{A}_i(0)=2.55\times10^{-3}$ in each case. 

\begin{figure}
\includegraphics[width=14.5cm]{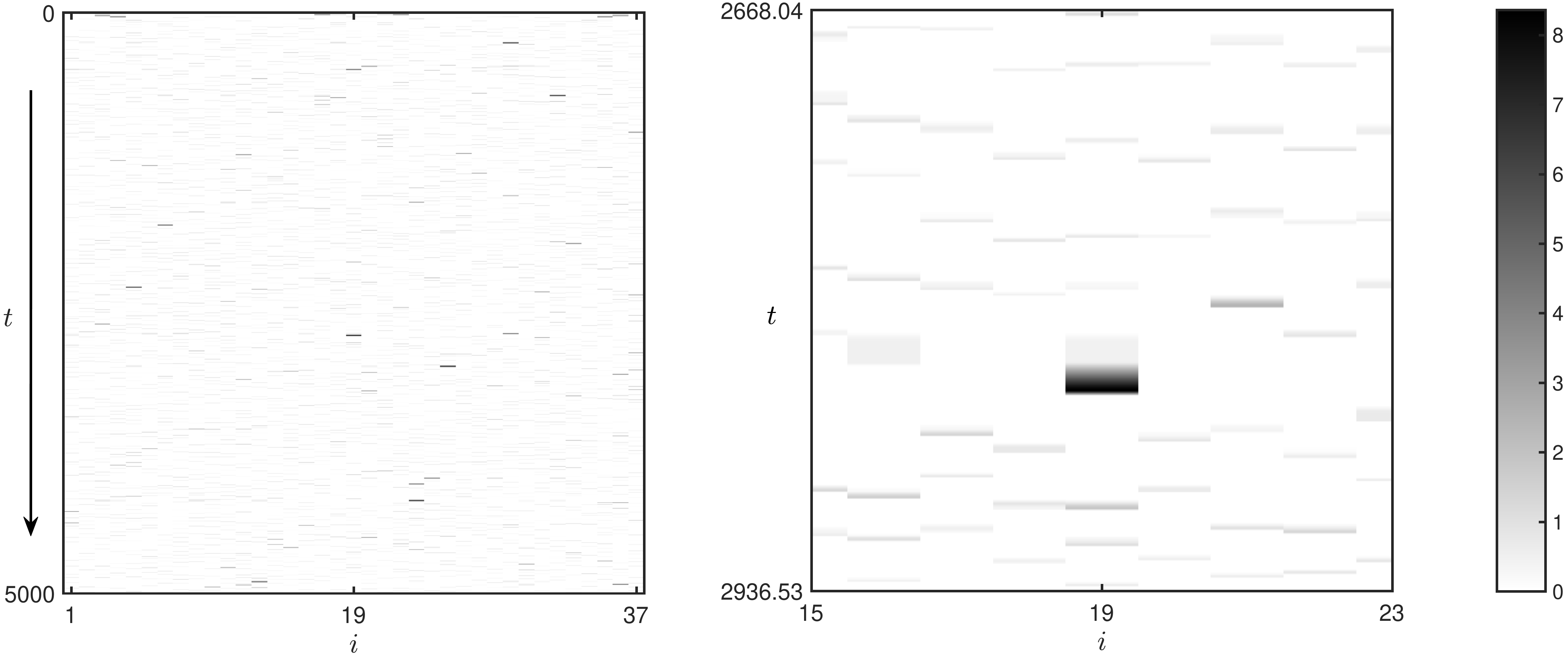}
\caption{Left panel: Space-time plot showing the evolution of $\log_{10}(\mathcal{A}_i)$ starting from a sine initial condition in a system of $N=37$ coupled oscillators, $1\le i\le 37$, with $K=2.1544\times10^{-6}$. Right panel: Zoom of the space-time plot to provide a close-up view of the largest amplitude
  event in the left panel.}
\label{fig2:evolexam}
\end{figure}

\begin{figure}
\includegraphics[width=7.5cm]{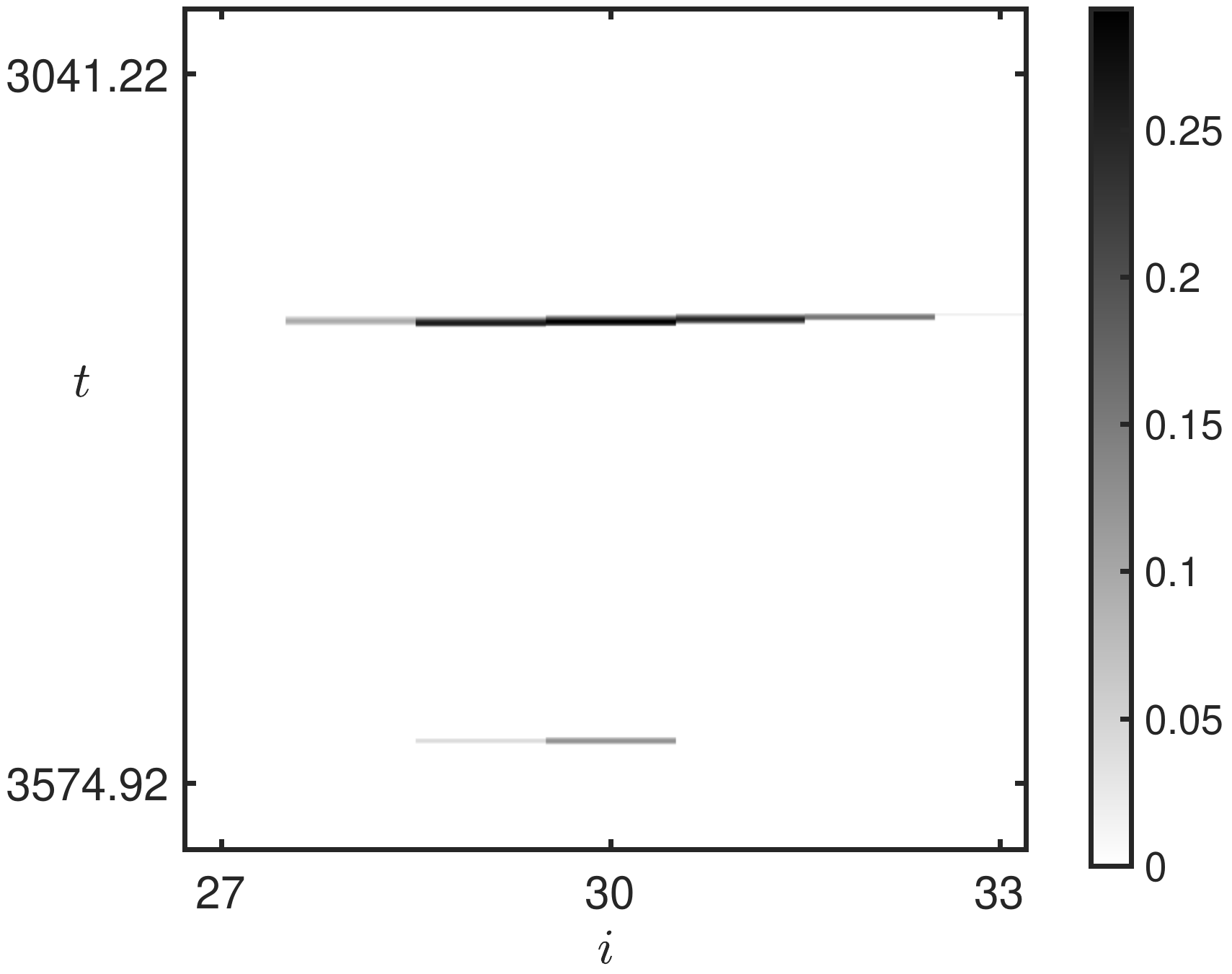}
\hspace{0.15cm}
\includegraphics[width=7.5cm]{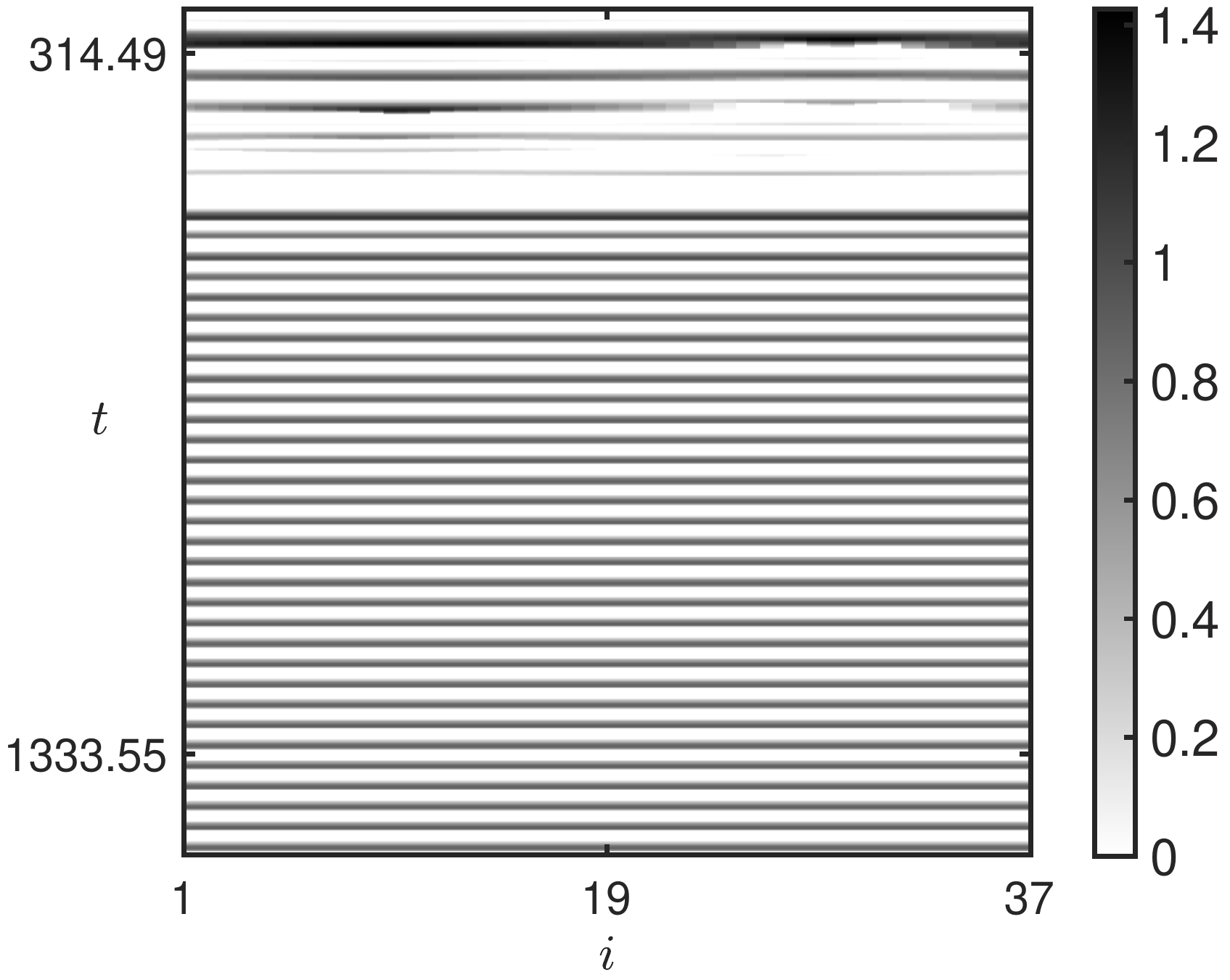}
\caption{Space-time plot showing the evolution of $\log_{10}(\mathcal{A}_i)$ starting from a sine initial condition in a system of $N=37$ coupled oscillators with $K=1 \times 10^{-4}$ (left panel), with maximum amplitude occurring at the $i=30$, and $K=2.2\times10^{-3}$ (right panel) with maximum amplitude occurring at $i=10$. The left panel shows that the large amplitude feature is not localized at a single oscillator and that neighboring oscillators respond coherently. In the right panel, all oscillators are synchronized and so oscillate in spatial coherence.}
\label{fig2:evolexam2}
\end{figure}

Figure\,\ref{fig2:evolexam} shows a space-time waterfall plot of the logarithm of the amplitude at each node, $\log_{10}\mathcal{A}_i$, for the case with coupling constant $K=2.1544\times 10^{-6}$ and starting from the sine initial condition. During this evolution, the range of variation in $\mathcal{A}_i$ spans eight orders of magnitude, reaching a maximum amplitude excursion of $\mathcal{A}_{19} = 2.28\times10^{8}$ at $t=2827.5$. The right panel zooms in close to the maximum rogue event to show the sharp and localized excursion in amplitude in more detail, thereby demonstrating that the lattice system is indeed capable of supporting rogue events. In particular we see that despite the initial sinusoidal variation of amplitude, this sinusoidal pattern is gradually disrupted, and we observe the emergence of multiple isolated large amplitude events, shown as black or dark gray events in Fig.\,\ref{fig2:evolexam}. The largest rogue event is shown in finer detail in the zoom in the right panel of Fig.\,\ref{fig2:evolexam}. The panels show that at this value of $K$ even immediate neighbors fail to synchronize with the rogue event nearby, as evidenced by the sharp spatial localization of the amplitude. In contrast, at the larger value $K=1\times 10^{-4}$ the neighbors do partially synchronize with the large amplitude event (Fig.\,\ref{fig2:evolexam2}, left panel) and all evolve to absolute values that are comparable to the maximal amplitude. The right panel of the figure shows that at the yet higher value $K=2.2\times 10^{-3}$, all the oscillators synchronize, leading to spatial coherence among all of them.

\begin{figure}
\centering{\includegraphics[width=10cm]{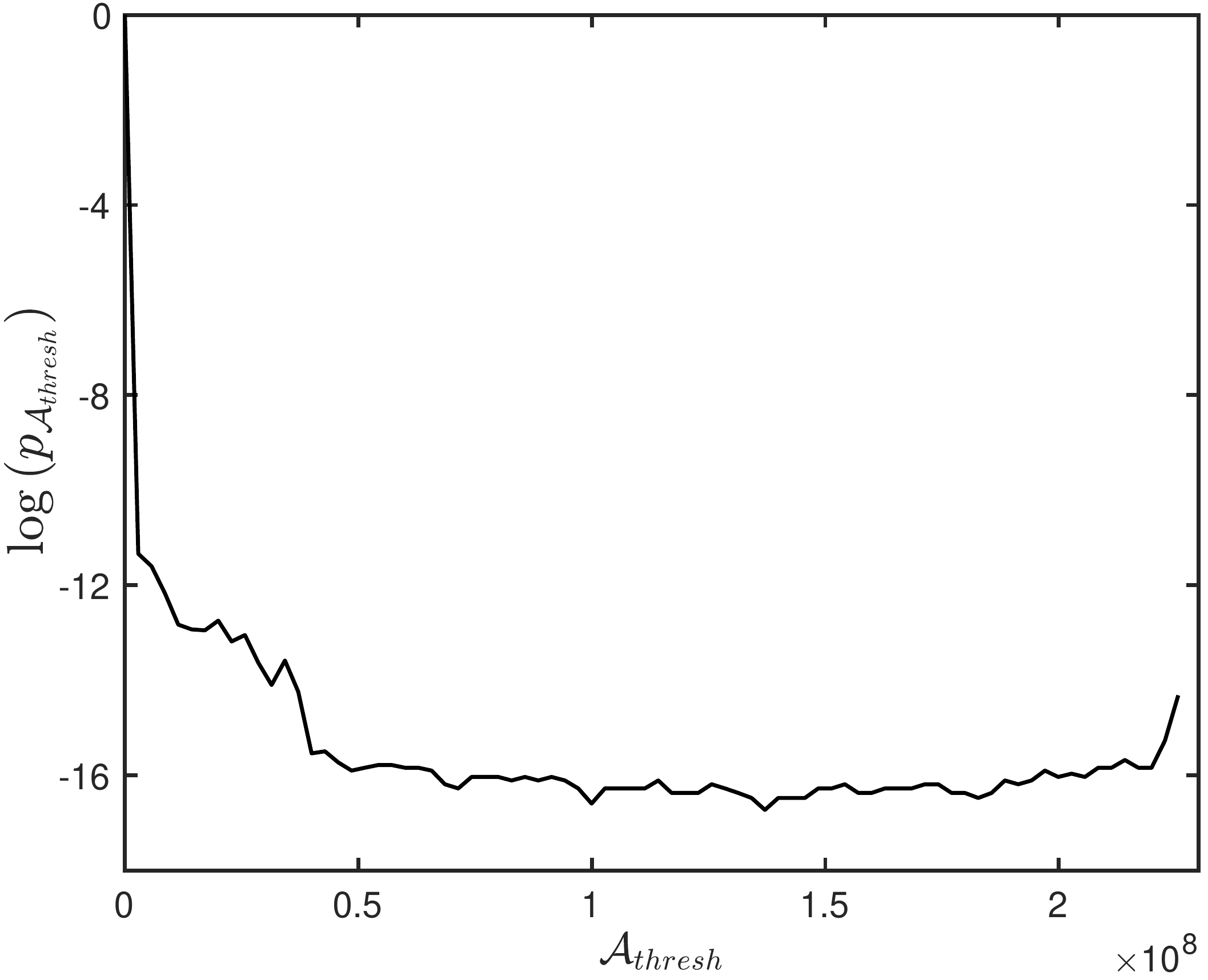}
}
\caption{Logarithm of the probability density function $p_{\mathcal{A}_{thresh}}$ for different amplitude thresholds $\mathcal{A}_{thresh}$ from the time evolution shown in Fig.\,\ref{fig2:evolexam}. We observe an increase in the probability of extremely large amplitudes, indicating that the occurrence of extreme events is more probable than what is expected from an exponential distribution of amplitudes.}
\label{fig3:pdf_k1em6}
\end{figure}

The observed large amplitude excursions develop and occur over a very short interval of time and do so irregularly both in space and time, hence their resemblance to a rogue event. In order to understand how frequent such excursions are, we compute the probability density function $p_{\mathcal{A}_{thresh}}$, which is calculated as the ratio of the number of instances when the amplitude falls between two thresholds, i.e., $\mathcal{A}_{thresh-1}<\mathcal{A}<\mathcal{A}_{thresh}$, to the total number of observations over the entire space-time run shown in Fig.\,\ref{fig2:evolexam}(a). In Fig.\,\ref{fig3:pdf_k1em6} we plot the logarithm of the probability density function $p_{\mathcal{A}_{thresh}}$ using $80$ bins spanning the range of amplitudes over the entire space-time run. From the figure we see that the occurrence of very high amplitude events is substantially higher than what would be expected if the distribution of events followed an exponential distribution. The emergent bimodality of the amplitude probability distribution with a second peak corresponding to extremely large amplitude events is representative of other albeit similar values of the coupling coefficient $K$ as well. It also persists in longer runs with the maximum peak amplitude reached gradually increasing as the run length increases.

We conclude that a diffusively coupled network formed of elements evolving according to Eqs.\,(\ref{eqn:temporalsys}) is capable of generating unpredictable, large amplitude excursions over a short time scale, whose occurrence is more probable than predicted via an exponential distribution of amplitudes. This system is therefore able to produce rogue  events. We next look at the effect of varying the coupling coefficient $K$ and describe the resulting changes to the space-time evolution in each case. 


\subsection{Effect of varying $K$}

\begin{figure}
\centering{\includegraphics[width=13cm]{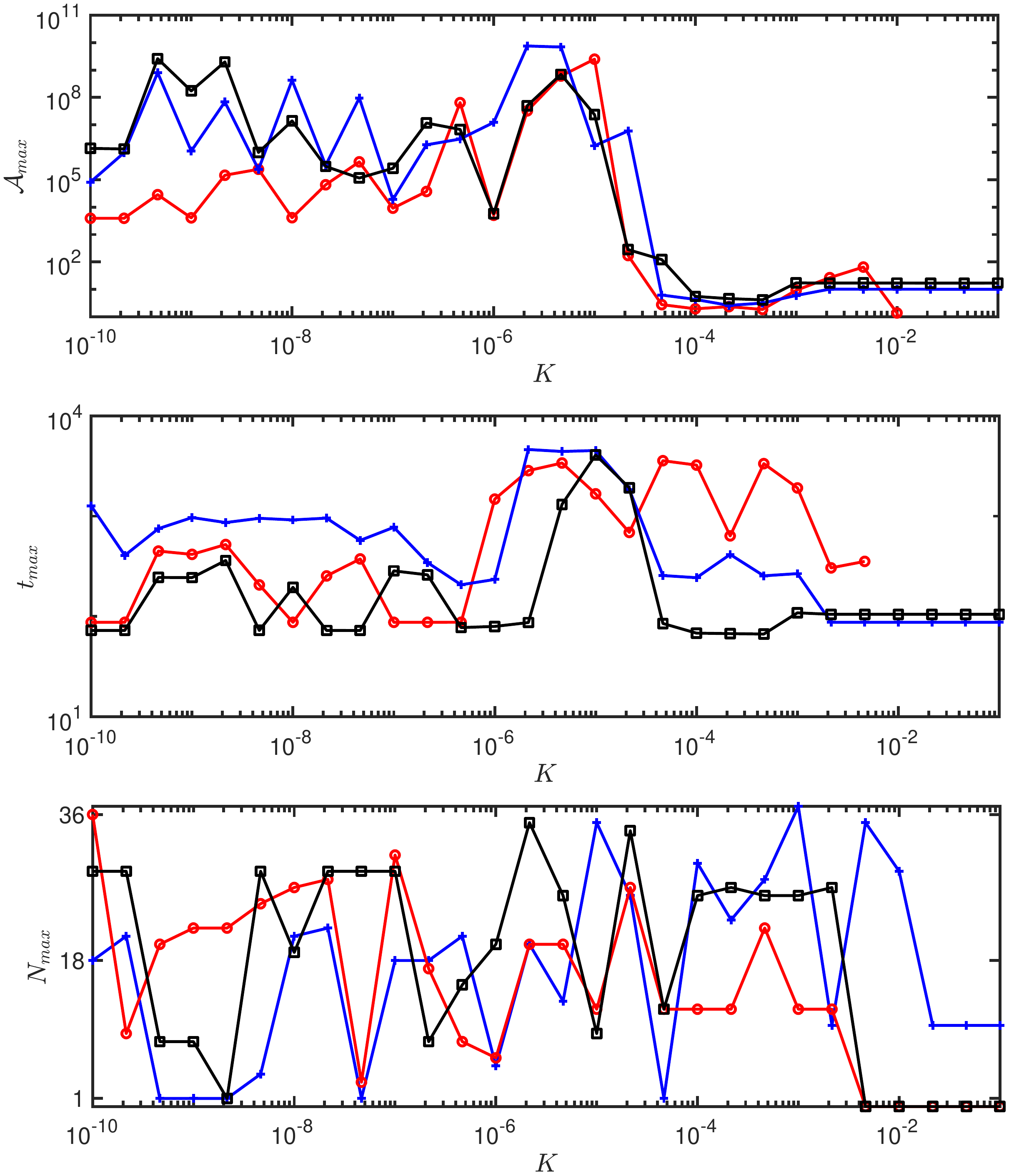}}
\caption{Red lines with circles show the results for single peak initial condition, blue lines with crosses show the results for the sine wave initial condition while black lines with square markers show results for the random initial condition. The top panel shows the variation in the maximum amplitude $\mathcal{A}_{max}$ (on log scale) as a function of the diffusion coefficient $K$. The middle panel shows the time $t_{max}$ taken to reach the maximal excitation, also on log scale, while the bottom panel shows the variation of the location $i=N_{max}$ of that maximal event among the individual oscillators.}
\label{fig4:ic_comparison}
\end{figure}

Having observed the model phenomenology for the special case of weak diffusive coupling ($K=2.1544\times10^{-6}$), we now turn to simulations over a range of diffusion coefficients, $10^{-10}\leq K\leq 1$, to appreciate the growing role of the coupling. The evolution resulting from the single peak initial condition is shown in Fig.~\ref{fig4:ic_comparison} as red lines with circles; that from the sine wave initial condition is shown as blue lines with crosses and, lastly, the one from the random initial condition is shown in black with square markers.

The top panel in the figure shows the maximum amplitude rogue wave observed over the evolution time, denoted $\mathcal{A}_{max}$, as a function of $K$, on a logarithmic scale. This maximum occurs at time $t=t_{max}$ at the $i=N_{max}$ node shown in the middle and bottom panels. We see that the maximum excitation remains large until $K\approx1\times10^{-5}$ after which the maximum amplitude abruptly decreases. Beyond this threshold, the diffusion coefficient is large enough to synchronize the nodes. For larger values of $K$, $\mathcal{A}_{max}$ increases again, albeit very slowly. For large enough values of $K$, we recover the regular periodic spiking behavior of a single oscillator for both the sine wave (blue line with crosses) and the random initial condition (black line with square markers). For the single peak initial condition (red line with circles) and large $K$ values, the diffusive coupling overcomes driving at each oscillator and the amplitude decays during subsequent evolution from the initial amplitude $\mathcal{A}_i(t=0)$ at every node $i$. For these evolutions, $\mathcal{A}_{max}$ corresponds to the initial amplitude, which is ${\mathcal O}(10^{-4})$.

The middle panel shows the time $t=t_{max}$ taken to reach the maximal excitation. For very low values of $K$ we observe that large amplitude excitations occur later in the case of a sine wave initial condition (compared to the other two initial conditions). At large values of $K$ (in the synchronized range), the largest amplitude excursions occur during transient evolution before all the oscillators reproduce the regular periodic oscillations expected from a single uncoupled oscillator. In the case of a single peak initial condition (red line with circles) $t_{max}=0$ once we reach the range of $K$ values where the initial amplitude decays with subsequent evolution, and this scenario is therefore not represented in the log-log plot of $t_{max}$ vs. $K$.

The bottom panel shows the location $i=N_{max}$ where the maximal excitation is observed as a function of $K$. We observe that irrespective of the initial condition, the location of the maximal amplitude is unpredictable; when $K$ is large the synchronized nature of the dynamics also leads to large scatter in $N_{max}$ but this scatter is no longer meaningful.

\begin{figure}
\includegraphics[width=13cm]{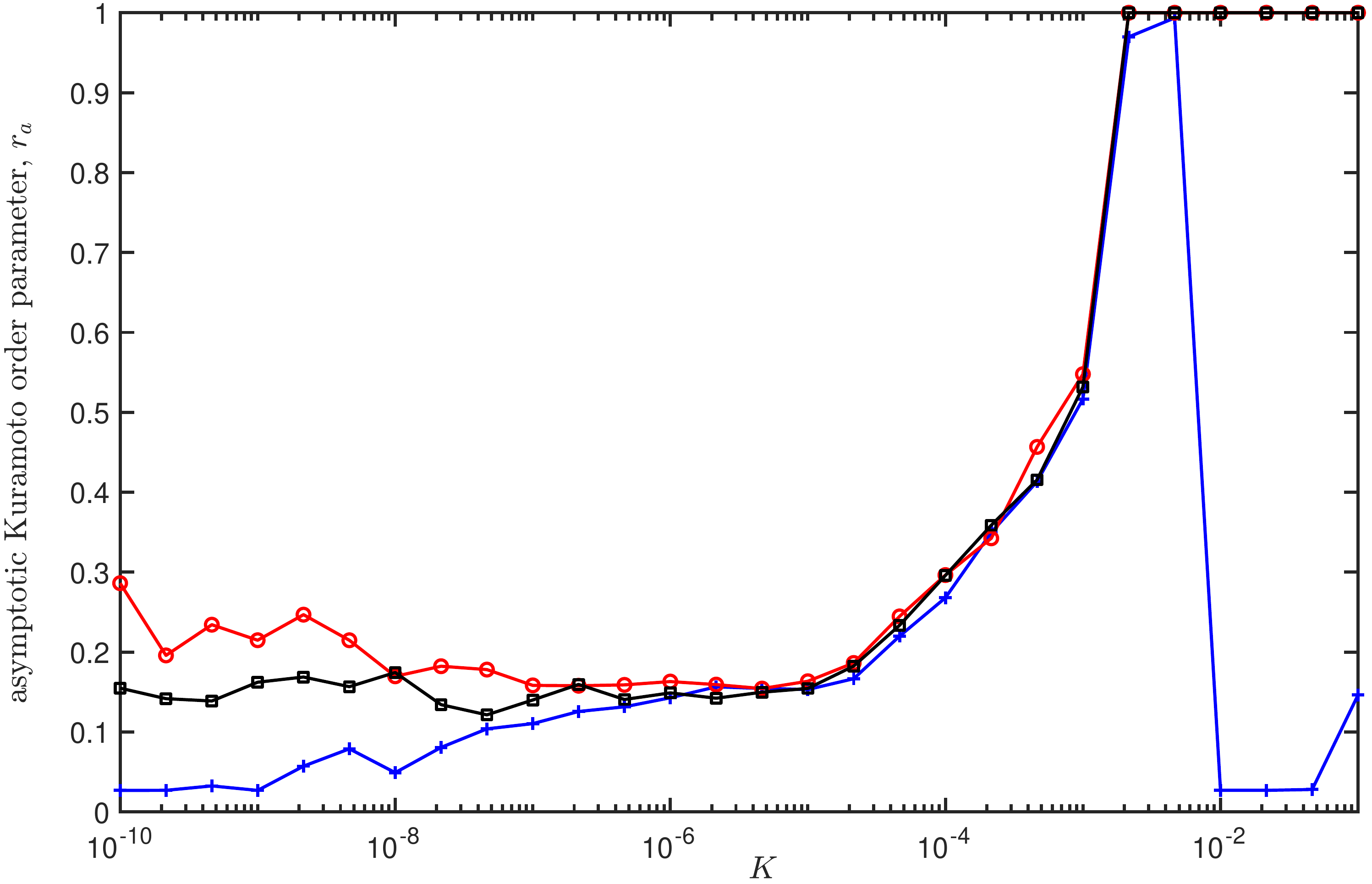}
\caption{Asymptotic value of the Kuramoto order parameter, $r_a = \lim_{t\rightarrow \infty} r(t)$, as a function of the coupling coefficient $K$ for different initial conditions: the case of a single peak initial condition (red lines with circles), a sine wave initial condition (blue lines with crosses) and a uniformly distributed random initial condition (black line with square markers).}
\label{fig4:kuramotor_K}
\end{figure}

In order to measure the effect of varying $K$ on synchronization between the nodes in this network, we compute the Kuramoto order parameter (discussed in detail in, e.g.,~\cite{STROGATZ20001}) for the space-time evolution at each value of the coupling parameter $K$. We measure the instantaneous degree of phase coherence using the quantity $r(t)$ defined by
\begin{equation}
    r(t) = \left| \frac{1}{N} \sum_{j=1}^{N} e^{i \theta_j} \right|\,,
\end{equation}
where $\theta_j = \tan^{-1} \left(\mathcal{I}m(z_{+})/\mathcal{R}e(z_{+})\right)$, and take the asymptotic value of $r$, $r_a  \equiv \lim_{t\rightarrow \infty} r(t)$, as representing the level of synchronization for the chosen level of diffusive coupling [$r(t)$ calculated with $z_{-}$ shows similar behavior at all $K$ values]. Figure \ref{fig4:kuramotor_K} shows the variation of $r_a$ as a function of $K$. As discussed in the introduction, at very low values of the coupling the nodes are uncoupled leading us to expect low values of $r_a$.
In fact, we observe three distinct values of $r_a$, depending on initial condition, for the following reasons. In the case of a single peak initial condition, most of the nodes in the network are `synchronized' at the zero value to begin with (leading to an accordingly larger initial value of $r_a$), while in a random initial condition there can be some nodes that start with similar values and in a sine wave initial condition there is instead a smooth variation of amplitude instead of multiple repeated values. 

In contrast, at large values of $K$, all nodes in the network may be fully synchronized, leading us to expect $r_a\approx 1$. This is indeed the behavior that we observe for the single-peak and random initial conditions at large values of $K$. However, for the sine initial condition with the smoothest variation of amplitude (shown in blue line with plus markers), we observe that at large values of $K$, $r_a$ falls sharply to very low values. We explain this as follows. Individual oscillators want to undergo regular periodic oscillations under the chosen parameter conditions. However, at large $K$, diffusion is strong enough to overcome such oscillations and causes the amplitude at every location in the ring to decay with time. In the asymptotic limit we then obtain small values for the Kuramoto order parameter $r_a$, indicative of effectively random phases computed from vanishingly small amplitudes. The final slight increase in $r_{a}\approx0.1$ is still within the range where we do not expect synchronized behavior.

\begin{figure}
\includegraphics[width=13cm]{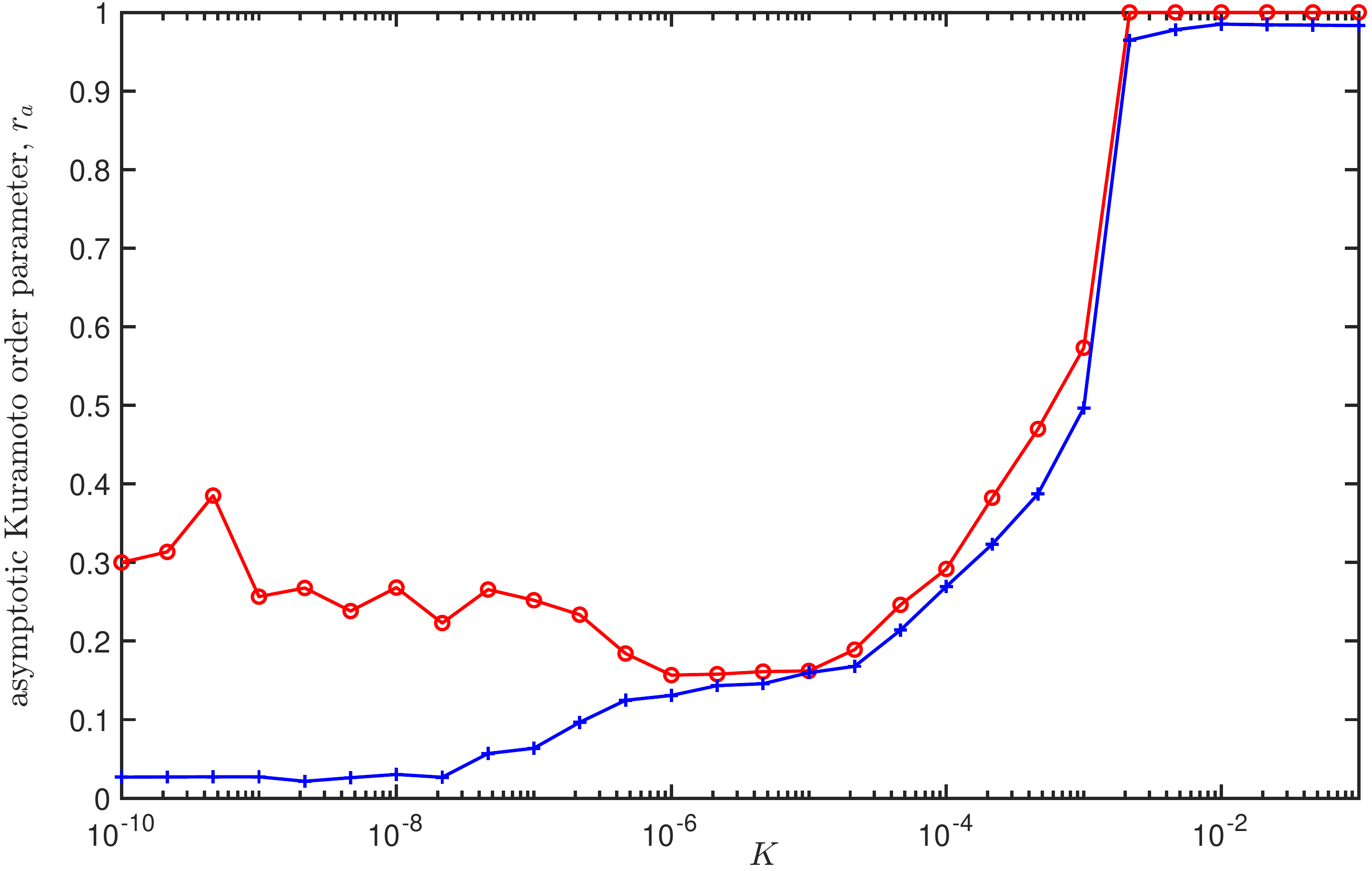}
\caption{Asymptotic value of the Kuramoto order parameter, $r_a = \lim_{t\rightarrow \infty} r(t)$, as a function of the coupling coefficient $K$ for the single peak initial condition (red lines with dots) and the sine wave initial condition (blue lines with crosses) starting with a larger initial amplitude distribution.}
\label{fig4:kuramotor_K_old}
\end{figure}

Of course, the observed behavior for increasing values of the coupling coefficient $K$ is also related to the initial amplitude distribution. In order to explore this dependence further, we show in Fig.\,\ref{fig4:kuramotor_K_old} a similar plot of the asymptotic Kuramoto order parameter $r_a$ as a function of $K$ for the single peak and the sine wave initial condition starting with a larger initial amplitude distribution. The Riemann sums of the $\mathcal{A}_i(t=0)$ for both initial conditions are again identical and the sine wave has a maximum value of $1\times10^{-2}$. We find that in both cases the evolution is able to reach fully synchronized behavior, $r_a\approx 1$, at large values of $K$. In this case, the sine wave initial condition is able to overcome diffusion at large $K$ values to retain the regular periodic oscillations at every oscillator in the ring resulting in large values of $r_a$. 

\subsection{Precursors for a high amplitude excursion}

As observed from the probability density function results (Figs.\,\ref{fig3:pdf_k1em6} and \ref{fig4:ic_comparison}) as well as the space-time plots (Fig.\,\ref{fig2:evolexam}), the location $N_{max}$ and the peak amplitude $\mathcal{A}_{max}$ cannot be directly predicted during the evolution of the coupled oscillator network. However, there is great interest in being able to identify precursors that can indicate impending large amplitude excursion in the form of a spatio-temporal rogue event. Previous work \cite{CousinsJFM2016} combined statistical analysis along with a nonlinear stability criterion for a local wave train to quantify the probability of the occurrence of a large amplitude event. These authors also identified a simpler precursor which only tracks the energy of the wave field within an identified critical length scale. Both these measures rely on the assumption that the basin boundary for a rogue  event is low-dimensional. A separate approach, that considers such waves as hydrodynamic instantons that can be analyzed within the framework of large deviation theory and computed via suitably tailored numerical methods, is explored in~\cite{grafke}. Since the dynamics of the oscillator ring is by construction related to the dynamics of a single oscillator, we opt here to leverage the known low-dimensional dynamics of a single oscillator to design a qualitative diagnostic that can identify impending large and rapid growth of amplitude at a given location in the network.

\begin{figure}
\includegraphics[width=12cm]{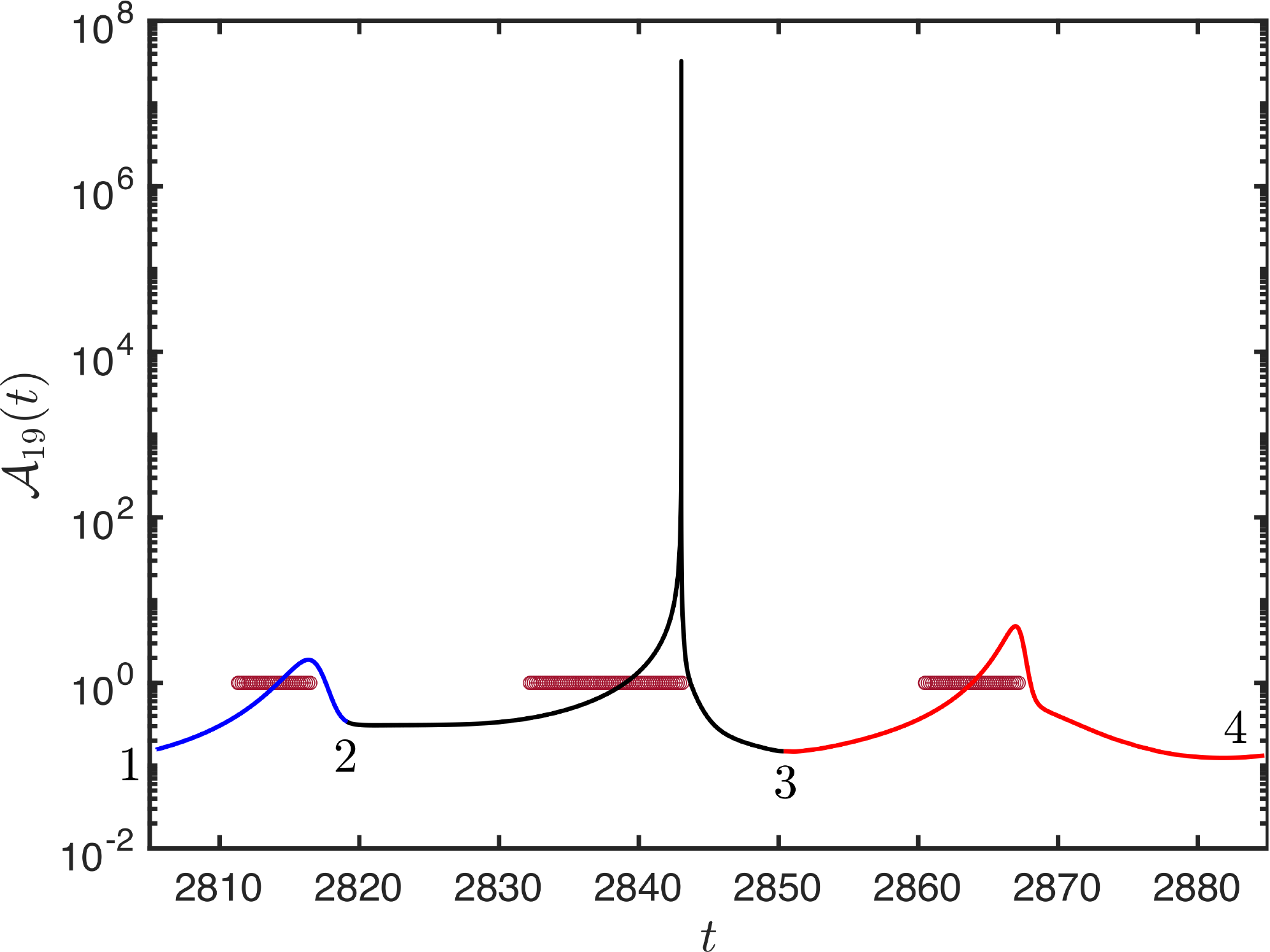}
\vspace{0.95cm}
\includegraphics[width=12cm]{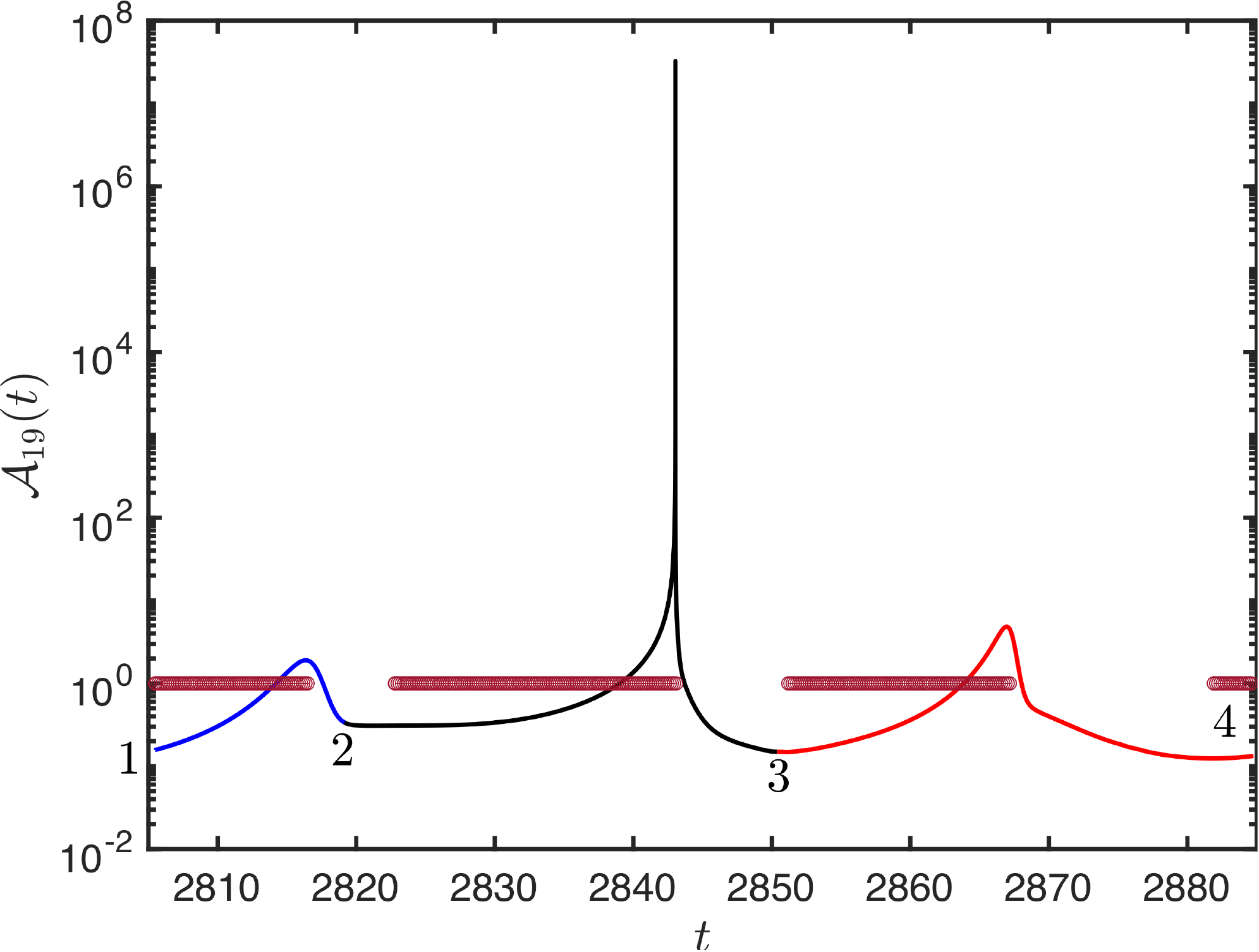}
\caption{Large amplitude events at $i=19$ as seen in the right panel of Fig.\,\ref{fig2:evolexam} in the time domain shown superposed on locations where the precursor $P$ identifies rapid growth (brown circles) due to the alignment of the trajectory with the stable direction of the saddle solution $u_{\infty}$ (top panel) and of the $v_{\infty}$ solution (bottom panel) of a single uncoupled oscillator.
} 
\label{fig6:event1wing}
\end{figure}

In Fig.~\ref{fig6:event1wing} we review the behavior during rogue events by plotting the amplitude $\mathcal{A}(t)$ of the $N_{max}=19$ oscillator in a semi-log plot for the evolution shown in the right panel of Fig.\,\ref{fig2:evolexam}. Each large amplitude excursion is noted in different colors: blue (between 1 and 2), black (between 2 and 3) and red (between numbers 3 and 4). 
Superimposed on this evolution and shown as brown circles are instances when our precursor $P$ (see the relevant definition below) indicates that the evolution is heading toward a large amplitude excursion. We observe that the qualitative precursor is able to identify each of the imminent large amplitude excursions well before the amplitudes have reached large values and independent of the amplitude at which the growth commences. In the rest of this section we detail how we construct this precursor.

The change of variables in Eq.\,(\ref{eqn:newvars}) allows us to identify the large amplitude events as $\mathcal{A} = \rho^{-1}$, where $\rho\ll1$. The limit of $\rho=0$ corresponds to the invariant subspace $\Sigma_{\infty}$, as this is the limit for which the amplitude $\mathcal{A}\rightarrow\infty$. In the following we denote the $u$ and $v$ solutions in $\Sigma_{\infty}$ with the subscript $\infty$. 

\begin{figure}
\includegraphics[width=12cm]{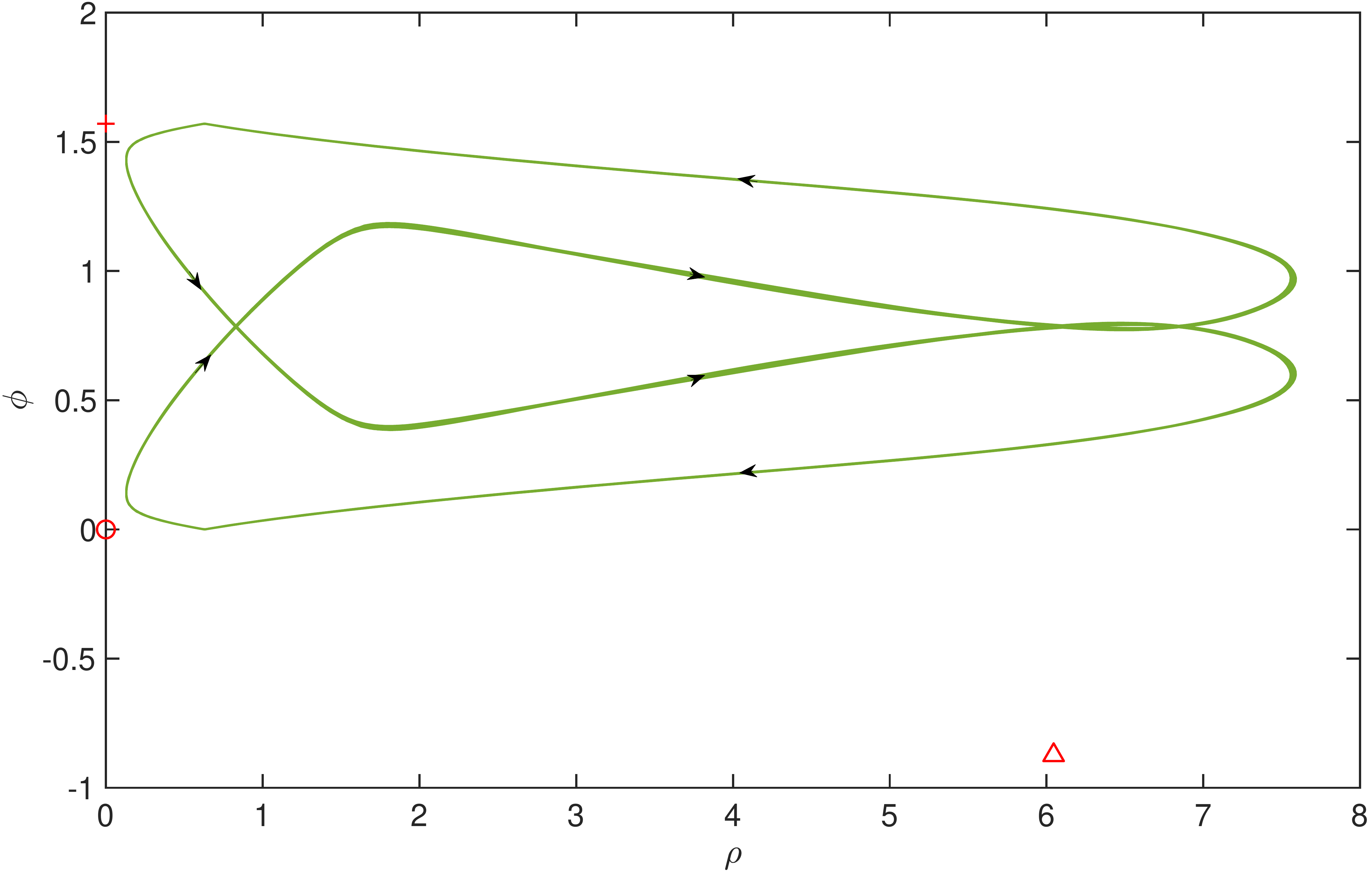}\\
\vspace{0.3cm}
\includegraphics[width=12cm]{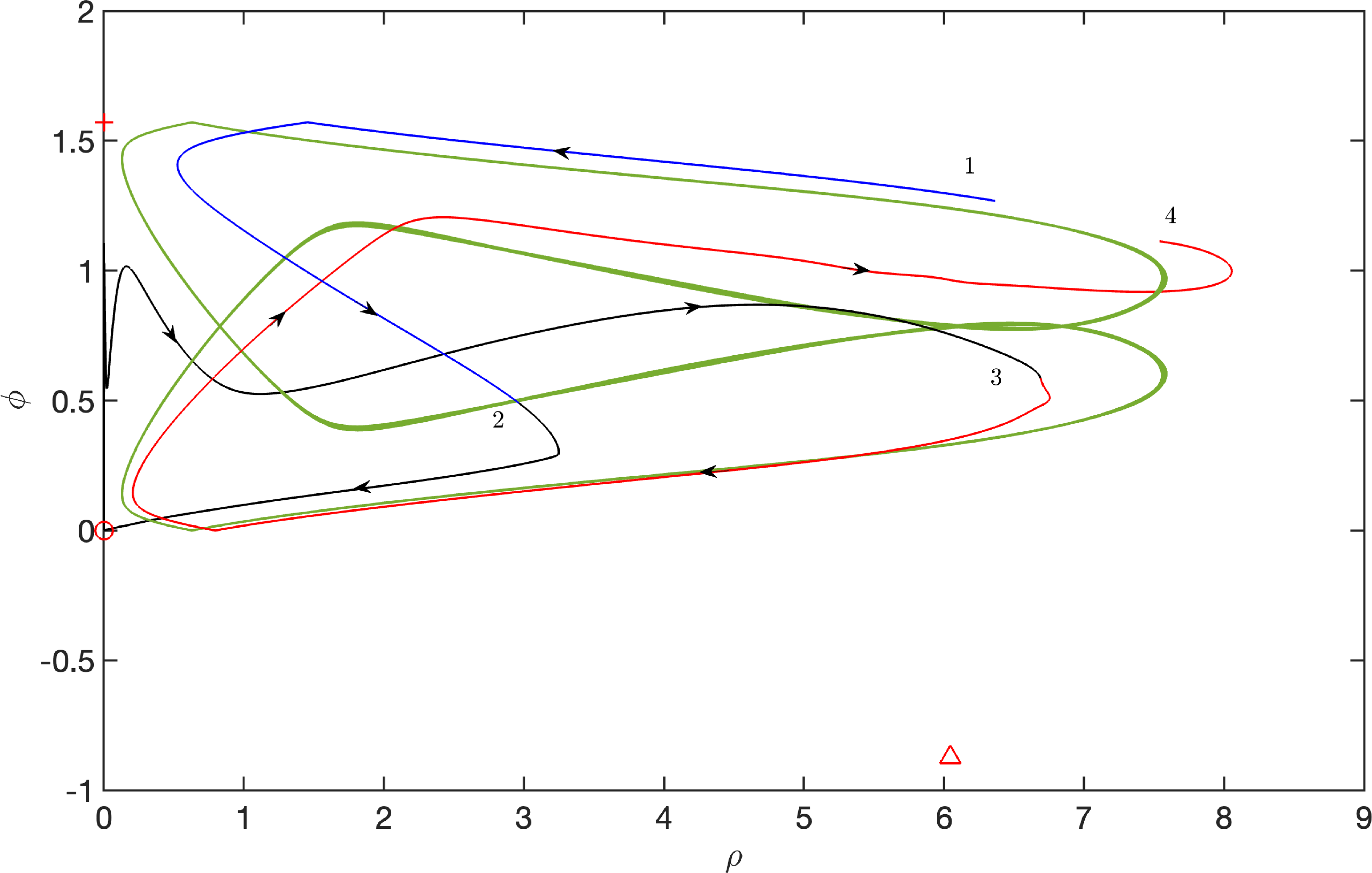}
\caption{Top panel: Phase portrait along the ($\rho$, $\phi$) plane during regular periodic spiking at a single node in the network with no coupling (in green line). Bottom panel: Same view in the ($\rho$, $\phi$) plane superposed with an evolution close to a large amplitude event with $K=2.1544\times 10^{-6}$ in the coupled network at $N=N_{max}$ (shown in black line). 
}
\label{fig5:rhothetaphi_ode_pde}
\end{figure}

Figure \ref{fig5:rhothetaphi_ode_pde} shows the trajectory (after an initial transient) of both the single uncoupled oscillator (in green, top panel) and the $i=19$ oscillator on a ring with $K=2.1544\times10^{-6}$ (in blue, black and red colors, bottom panel) projected onto the variables $\rho$ and $\phi$. In this figure, red markers (pluses, circles and triangles) in the panels indicate the locations of the invariant solutions for an uncoupled oscillator in the ($\rho,\theta,\phi$) variables. Circles indicate solutions of the so-called $u_{\infty}$ type~\cite{PhysRevLett.80.5329,MOEHLIS2000263}, pluses indicate solutions of $v_{\infty}$ type and triangles indicate finite amplitude solutions that are mixed $u/v$ states. These three solutions represent fixed points of the ($\rho$,$\theta$,$\phi$) system \ref{eqn:rhothetaphi} and represent traveling states of the underlying individual oscillator dynamics described by (\ref{eqn:temporalsys}) with the chosen level of asymmetry at the parameters given in Eq.\,(\ref{eqn:odeparam}).

For this oscillator, we calculate the eigenvalues of each solution in the $\Sigma_{\infty}$ subspace \cite{MOEHLIS2000263} and find that the point $u_{\infty}$ at ($\rho$,$\theta$,$\phi$) = ($0$,$\pi/2$,$0$) is a saddle [red circle in Fig.\,\ref{fig5:rhothetaphi_ode_pde}(a)] while $v_{\infty}$ at ($\rho$,$\theta$,$\phi$) = ($0$,$\pi/2$,$\pi/2$) is an unstable spiral [red plus in Fig.\,\ref{fig5:rhothetaphi_ode_pde}(a)]. Integration of the Eqs.~(\ref{eqn:rhothetaphi}) reveals a stable periodic spiking orbit (green line in panel (a), black arrows indicating direction of forward time) that periodically approaches both $u_{\infty}$ and $v_{\infty}$, i.e., small values of $\rho$. Both of these excursions correspond to amplitude spikes, with no discernible difference between them. We expect the dynamics of a diffusively coupled ring of such nodes to follow this orbit at small values of diffusive coupling $K$. Indeed, when all the oscillators in the ring are initialized with this same periodic state and the same temporal phase, the coupled evolution on the ring retains this synchronization. This means that the rogue event-bearing state observed for the ring of oscillators coexists with a regular periodic oscillation reflecting the dynamics of a single uncoupled oscillator.

In Fig.\,\ref{fig5:rhothetaphi_ode_pde}(b), we superimpose the evolution of the $i=19$ node in the diffusively coupled ring shown previously in Fig.\,\ref{fig6:event1wing} on the dynamics of the uncoupled system (panel (a), green trajectory), both projected to the ($\rho$,$\phi$) plane. The blue, black and red parts of the trajectory match those in Fig.\,\ref{fig6:event1wing}. Between points $1$ and $2$ (segment shown in blue) the solution approaches $v_{\infty}$ (red plus). This approach is responsible for the first spike in Fig.\,\ref{fig6:event1wing}. Beyond point 2 (segment shown in black) the trajectory is able to return to extremely small values of $\rho$, $\rho = 4.39\times10^{-9}$, but this time due to an approach to $u_{\infty}$ (red circle). This close approach is responsible for the very large amplitude spike in Fig.\,\ref{fig6:event1wing}. Beyond point $3$ (segment shown in red), the trajectory returns to $u_{\infty}$ but does not reach such small values of $\rho$. This excursion is responsible for the third spike in Fig.\,\ref{fig6:event1wing}. Thus the evolution of the $i=19$ oscillator recapitulates the dynamics of a single oscillator, but does so irregularly and with occasional excursions close to the $\rho=0$ fixed points, resulting in a large amplitude spike, i.e., a rogue wave. This comparison also suggests that the coupling to nearest neighbors may occasionally lead to smaller values of $\rho$ (i.e., bigger spikes) as in the segment from $2$ to $3$, but also to larger values of $\rho$, as in the segment from $3$ to $4$.

Given that the uncoupled dynamics has a pair of saddles in the $\Sigma_{\infty}$ subspace, we can expect a large amplitude excursion if the evolution is aligned with the stable direction of the saddle $u_{\infty}$, say, which we will call as $\bar{V}_1$. Further, we expect the amplitude to continue to grow until the projection of the current state on the fastest unstable eigendirection of the saddle $u_{\infty}$, which we will call $\bar{V}_2$, starts to increase. We use the above notions to design a qualitative precursor for a large amplitude excursion as follows: 
 \begin{equation}
P = \begin{cases}
1, & \textrm{if $\mathcal{P}(state,\bar{V}_1) >0$ and $\mathcal{P}(state,\bar{V}_2)$ is decreasing in time}
\\
0, & \textrm{otherwise.}
\end{cases}
 \end{equation}
Here $\mathcal{P}$ is the projection of the current state along the respective eigendirections $\bar{V}_1$ or $\bar{V}_2$. With the above definition, we identify instances in time where the local dynamics is aligned close to the attracting direction of the saddle $u_{\infty}$ and is not evolving along the unstable direction of $u_{\infty}$. We determine the temporal variation of the projection of the current state along $\bar{V}_2$ via a simple first order approximation of the derivative. When both these conditions are satisfied and ${P}=1$, we expect the dynamics to continue to evolve along the stable direction $\bar{V}_1$ of the saddle $u_{\infty}$, implying that $\rho\rightarrow0$ and the amplitude therefore grows. This condition is what we identify as a qualitative precursor of an impending large amplitude excursion. When these conditions are not satisfied, we have ${P}=0$ and we do not expect to see a large amplitude event. 

As already mentioned, overlaid on the amplitude evolution in Fig.\,\ref{fig6:event1wing}(a) and shown in brown circles, are time instances where the above criterion predicts a precursor event ($P=1$), indicating that a large amplitude excursion is imminent. We see that this diagnostic is able to identify the growth intervals of all three amplitude excursions present here. A similar definition of a precursor can also be created with respect to the eigendirections of the $v_{\infty}$ solution (Fig.\,\ref{fig6:event1wing}(b)). Thus, estimating to what extent the current state of a node in a coupled ring of oscillators maps on the dynamics of a single uncoupled oscillator allows us to identify impending large amplitude events in this system. We believe that this type of diagnostic could be relevant to other systems, provided that a mathematical characterization of the large amplitude solutions of a single uncoupled node is available.

\subsection{Amplitude-dependent diffusive coupling}
 
 Having examined the case of uniform coupling across the nodes of our ring, we now wish to explore the potential impact of heterogeneity in the lattice coupling, cf.~\cite{Braiman1995,Sugitani2021,Zhang2021}. Specifically, we suppose that the spatial coupling depends on the current value of $z_{\pm}$ via the amplitude $\mathcal{A}$ according to
\begin{equation}
K_i = K_0 \mathcal{A}_i(t) = K_0 (|z_{i+}|^2 + |z_{i-}|^2),\label{ampcoupling}
\end{equation}
with $K_0$ being a tunable parameter.
 
 \begin{figure}
\centering{\includegraphics[width=12cm]{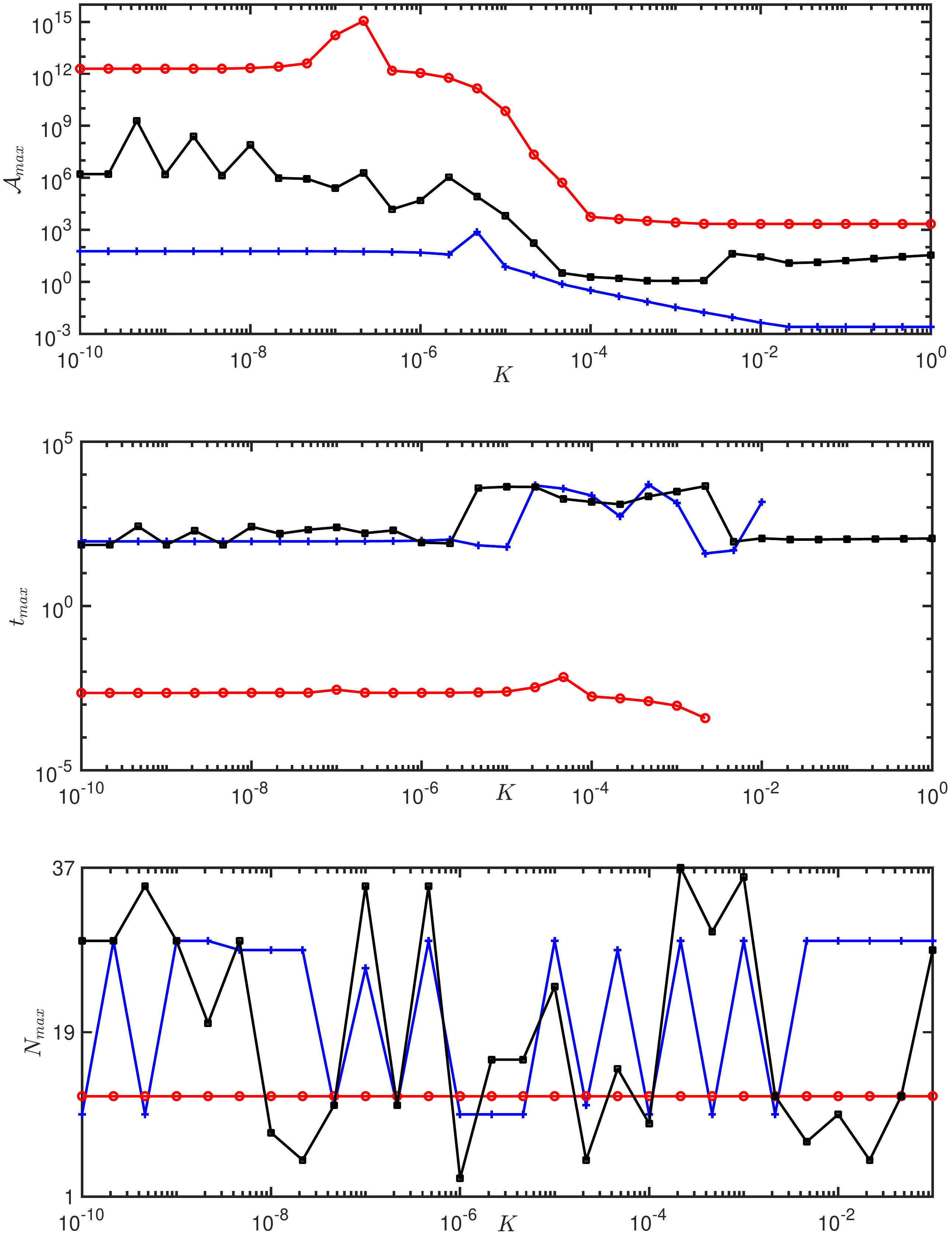}}
\caption{Same as Fig.~\ref{fig4:ic_comparison} but for the amplitude-dependent coupling (\ref{ampcoupling}). Red lines with circle markers show results for the single peak initial condition while blue lines with $+$ markers are the results for the sine wave initial condition. The top panel shows the variation in the maximum amplitude of observed excitations (on log scale) as a function of the coupling coefficient $K_0$. The middle panel shows the variation in the time taken to reach the maximal excitation $t_{max}$ on log scale, also as a function of $K_0$. Finally, the bottom panel shows the variation of the location $N_{max}$ of the maximal amplitude among the individual oscillators as a function of $K_0$. Note that the single peak initial conditions were initialised at $N=12$. }
\label{fig7:ic_comparison_kdepr}
\end{figure}

 Here we observe some qualitative differences in the behavior of the system for different initial conditions. As shown in Fig.\,\ref{fig7:ic_comparison_kdepr}, for the single peak initial condition (the red lines with circle markers) the maximal events observed are much larger, occur much earlier in the evolution (compared to other initial conditions) and are all concentrated at the location of the initial peak. This is expected, as the vast majority of the lattice nodes is initialized with near-vanishing amplitude thereby preventing the outward propagation of the initial disturbance even for $K_0\ne0$. In contrast, random and sine distributed initial conditions are observed to yield excitations with lower amplitudes over the whole range of $K_0$ considered.  For a sine wave initial condition, these mostly occur at the locations of the peaks in the sine wave initial condition while there is no preferential location for the random initial condition. As expected, the single peak initial condition can only exhibit rogue events at the location that is initialized with nonzero amplitude.






\section{Conclusions and Future Challenges}

In this work we have revisited a system that is known to exhibit bursting over a coherent region of space, and
reformulated the problem to permit the activation of spatial degrees of freedom. We found that the resulting
system could indeed generate extreme events that were localized in both time and space, and that occurred
more-or-less at random locations on a periodic ring of such bursters and at random times. Our system offers
an intriguing alternative to more conventional studies of rogue wave formation based on integrable Hamiltonian
partial differential equations, typically the nonlinear Schr\"odinger equation and its variants. The latter
approach has met with considerable success, and there is good evidence that rogue waves resembling the
Peregrine soliton~\cite{Peregrine1983} and its periodic and higher order generalizations do in fact occur
in wave experiments in a channel geometry (see, e.g.,~\cite{Chabchoub2012}). Our aim has been to propose an
alternative mechanism that could give rise to such extreme events in distributed forced dissipative lattice systems.
The proposed mechanism is fundamentally based on a strong resonance between two almost degenerate modes 
and has the remarkable property that it permits  excitations of {\it arbitrarily} large amplitude. The model
has the welcome additional property that its dynamics ``at infinity'' is well understood. We considered
a nonlinear dynamical lattice consisting of diffusively coupled elements of the above type, and demonstrated
that such lattices can manifest a phenomenon resembling rogue events, i.e., waves that ``appear out
of nowhere and disappear without a trace''~\cite{AKHMEDIEV2009675}. We showed in particular that
such events may be present even when the individual oscillators oscillate periodically, and explained how
this behavior depends on the (weak) diffusive coupling between the oscillating elements. In addition, we
demonstrated the possibility of synchronization at larger coupling strength, quantified the distribution of
the rogue amplitudes in terms of a bimodal probability distribution and examined the synchronization
properties of the system using Kuramoto-type order parameter diagnostics. Importantly,  we also presented
an approach that enabled us to predict reliably impending rogue events through a quantitative understanding
of (infinite amplitude) solutions and their eigenvector characteristics. We believe that this approach is
suitable for implementing machine learning techniques, and will explore this approach in a future publication.

Our emphasis on strong resonance between nearly degenerate modes differs fundamentally from alternative
approaches based on the nonlinear evolution of modulational instabilities but connects the rogue wave phenomenon
to the dynamics of systems exhibiting large amplitude sloshing \cite{Zeff2000,Faltinsen2009} and our system is
arguably one the simplest ones of this type. We have leveraged the behavior of coupled oscillators with approximate
1:1 temporal resonance \cite{Steen1982} but incorporated in our approach the possibility that standing oscillations
are themselves unstable to traveling modes, cf.~\cite{Crawford1991}. It is ultimately this destabilization of the
standing mode that permits the large amplitude bursting behavior present in our model. The spatial coupling of
our bursting elements is designed to activate scales larger than the length scale of each element and hence capture
the dynamics of large scale systems where similar destabilization is present \cite{Batiste2001}.

Naturally, there exist numerous directions for further study. While here we have given a proof-of-principle of
localized dynamics in space-time, it does not escape us that the original ODE system in Eqs.~(\ref{eqn:temporalsys})
possesses a substantial wealth of additional possible states as the relevant parameters
vary~\cite{PhysRevLett.80.5329,MOEHLIS2000263}. In this light, a further study how such additional fixed points
(especially so in the invariant subspace at infinite amplitude) may affect the dynamics of a diffusively coupled
lattice system is certainly merited, as is a study of the effect of random coupling strengths between adjacent nodes,
be these quenched or stochastically varying in time \cite{Braiman1995,Sugitani2021,Zhang2021}. Moreover, the mechanism
of extreme event production put forth herein is not restricted to one-dimensional lattices (as is often the case for
integrable Hamiltonian systems) but generalizes naturally to higher dimensions, a topic also worth exploring in its
own right. Such studies are currently in progress and will also be reported in a future publication. 

\vspace{5mm}

{\it Acknowledgments.} PGK gratefully acknowledges discussions with
Ji Hun (Jimmy) Hwang whose Honors Thesis provided some early-stage
dynamical simulations for the lattice system considered herein.
This material is based upon work supported by the US National Science
Foundation under Grants No. DMS-1908891 (EK) and PHY-2110030 (PGK).
EK and PGK also gratefully acknowledge Professor I.G. Kevrekidis
for initiating their discussions on this topic.


\section{APPENDIX: the ($\rho,\theta,\phi$) system}

We consider the ($\rho$,$\theta$,$\phi$) formulation from \cite{MOEHLIS2000263} in terms of a rescaled time $d\tau/dt = 1/\rho$. In this formulation Eqs.\,(\ref{eqn:temporalsys}) become
\begin{align}
\frac{d \rho}{d \tau} &= \rho(2 A_R + B_R(1+\cos^2\theta) + C_R \sin^2\theta\cos2\phi) - 2 (\lambda + \triangle \lambda \cos\theta) \rho^2\,,\nonumber\\
\frac{d\theta}{d\tau} &= \sin\theta(\cos\theta(-B_R +C_R\cos2\phi) - C_I \sin2\phi) - 2\triangle \lambda \rho \sin\theta\,,\nonumber \\
\frac{d\phi}{d\tau} &= \cos\theta (B_I - C_I \cos2\phi) - C_R \sin2\phi + 2\triangle \omega \rho\,,\nonumber\\
\label{eqn:rhothetaphi}
\end{align}
where the subscripts $R$, $I$ indicate real and imaginary parts.
In order to determine equilibria for this set of equations, we recast these equations in terms of the state vector 
$$X = (X_1,X_2,X_3,X_4,X_5) = (\rho,\cos\theta,\sin\theta,\cos2\phi,\sin2\phi)\,.$$
With this new state vector, the dynamics governing the system constitutes a system with constant coefficients, which implies that its equilibria can be determined by solving the associated set of polynomial equations
\begin{align}
-X_1 [ 2 A_R + B_R (1+X_2^2) + C_R X_3^2 X_4] - 2 (\lambda + \triangle \lambda X_2)X_1^2 &=0 \nonumber \\
X_3 [X_2(-B_R +C_R X_4) - C_I X_5] - 2 \triangle \lambda X_1 X_3 &=0 \nonumber \\
X_2(B_I - C_I X_4) - C_R X_5 + 2 \triangle \omega X_1 &= 0 \nonumber\\
X_2^2 + X_3^2 - 1 &= 0 \nonumber \\
X_4^2 + X_5^2 - 1 &= 0 \,.
\end{align}
Here the last two equations arise from the conditions that must be satisfied by the transformation of the sine and cosine functions into the new variables. In this recast form, we have a fully determined system of polynomial equations for the five unknowns and we use homotopy methods to determine all real, finite and nontrivial solutions of the resulting system using Bertini \cite{bertinibook}. 

\bibliography{rogue_wave7}

\end{document}